\documentclass[grl]{AGUTeX}

 \usepackage[dvips]{graphicx}
 
 \setkeys{Gin}{draft=false}
\authorrunninghead{WESTBROOK AND ILLINGWORTH}

\titlerunninghead{DOPPLER LIDAR MEASUREMENTS IN CIRRUS}

\authoraddr{C. D. Westbrook,
Department of Meteorology, University of Reading, Berkshire, UK.
(c.d.westbrook@reading.ac.uk)}

\begin{document}


%
%

\title{Testing the influence of small crystals on ice size spectra\\ using Doppler lidar observations}
%

%
%


\author{C. D. Westbrook and A. J. Illingworth}
\affil{Department of Meteorology, University of Reading, Berkshire, UK.}



%
%
%

%
%


\begin{abstract}
Measurements of the vertical velocity of ice crystals using a $1.5\mu\mathrm{m}$ Doppler lidar are described. The statistics from a continuous sample of stratiform ice clouds over 17 months are analysed: the distribution of velocity varies strongly with temperature, with average Doppler velocities of $\approx0.2\mathrm{ms^{-1}}$ at $-40^{\circ}\mathrm{C}$ increasing to $\approx0.6\mathrm{ms^{-1}}$ at $-10^{\circ}\mathrm{C}$ presumably due to particle growth and broadening of the size spectrum. We examine the likely influence of small crystals less than $60\mu\mathrm{m}$ by forward modelling their effect on the area-weighted fall speed, and comparing the results to the lidar observations. The comparison strongly suggests that the concentration of these small crystals in most ice clouds is much lower than measured in-situ by cloud droplet probes. The discrepancy is attributed to shattering of large crystals on the probe inlet, and we argue that these numerous small particles should not be included in numerical weather and climate model parametrizations.
\end{abstract}

%
%

%

\begin{article}

%
%

\section{Introduction}
There has been much controversy over the number of small sub-$60\mu\mathrm{m}$ particles in ice clouds. Measurements from forward-scattering cloud droplet probes suggest that these tiny particles are present in very large concentrations (eg. Boudala \textit{et al} 2002, Platt \textit{et al} 1989, Gayet \textit{et al} 2007, Lawson \textit{et al} 2006; see also Heymsfield and McFarquhar 2002), with important implications for the cloud's microphysical evolution, its interaction with radiation (Boudala \textit{et al} 2007, De Leon and Haigh 2007, Zender and Kiehl 1994), and remote sensing (Donovan 2003). On the other hand, there is evidence (Gardiner and Hallett 1985, Field \textit{et al} 2003, McFarquhar \textit{et al} 2007, Heymsfield 2007) which suggests that these small particles are in fact instrument artefacts, caused by large crystals shattering on the probe inlet, with the detector then sampling the numerous tiny fragments. 

General circulation models (GCMs) typically represent ice particle size spectra using a simple exponential or gamma distribution (eg. Wilson and Ballard 1998, Mitchell 1991), with the concentration of particles  of diameter $D$ given by 
\begin{equation}
N(D)=N_0 D^{\mu}\exp(-\lambda D).
\label{gamma}
\end{equation}
The parameters $N_0$ and $\lambda$ are either diagnosed as a function of temperature $T$ or deduced from the model ice water content (IWC), whilst the parameter $\mu$ is typically assumed to take a constant value (zero for a pure exponential). However, if the measured concentrations of small crystals are genuine, equation \ref{gamma} substantially under-represents their numbers. 

\begin{figure}
\noindent\includegraphics[width=20pc]{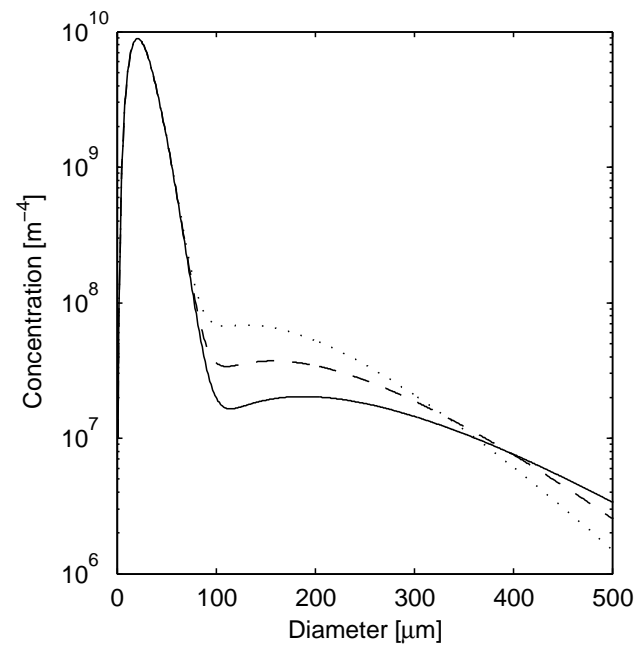}
\caption{\label{bimodal}Example bimodal size spectra for $T=-15^{\circ}\mathrm{C}$ (solid line), $-25^{\circ}\mathrm{C}$ (dashed line), and $-35^{\circ}\mathrm{C}$ (dotted line). IWC is fixed at $0.01\mathrm{gm^{-3}}$.}
\end{figure}

To better represent the small crystal observations, Ivanova \textit{et al} (2001) have taken cloud droplet probe and 2D optical probe measurements from 17 flights through stratiform ice clouds (a total of over 1000 individual spectra), and used them to parametrize a two-mode size distribution for input into GCMs. In this scheme, particles larger than $\approx100\mu\mathrm{m}$ are represented by the simple gamma distribution described above, with $\lambda$ diagnosed from the cloud temperature. However, a tall, narrow bump centred at around $25\mu\mathrm{m}$ is also added to represent the small particles seen by the cloud droplet probe. Examples of these parametrized spectra at temperatures of $-15$, $-25$ and $-35^{\circ}\mathrm{C}$ are shown in figure \ref{bimodal} for a fixed IWC of $0.01\mathrm{gm^{-3}}$. Concentrations of crystals $25\mu\mathrm{m}$ in diameter are typically 2-3 orders of magnitude higher than crystals $250\mu\mathrm{m}$ in diameter at these temperatures, and will strongly influence the lower moments of the particle size distribution (eg. optical extinction, deposition/evaporation rates). A very similar representation of the size spectra was suggested for remote sensing studies by Donovan (2003).

Mitchell \textit{et al} (2008) applied the two-mode parametrization outlined above to a GCM and showed that these tiny, barely-falling crystals would have a significant effect on ice cloud coverage and radiative forcing, particularly in cold tropical clouds. It is therefore vital to determine whether these concentrations of small crystals are genuine or not, in order that simulation of ice cloud in numerical weather and climate models is to be realistic.

Westbrook (2008) suggested that Doppler lidar measurements could be a sensitive test for the presence of genuine populations of small crystals, since the Doppler velocity would be significantly weighted by the small mode (which falls at only a few $\mathrm{cm~s^{-1}}$). In what follows, we test this idea, using 17 months of continuous observations from the Doppler lidar based at the Chilbolton Observatory in Hampshire, UK, and compare the results to a forward model of Ivanova \textit{et al}'s parametrization of the ice particle size spectra, both with and without the controversial small mode. 

\section{Doppler Lidar observations}
The measurements were made using a $1.5\mu\mathrm{m}$ Doppler lidar (HALO Photonics Ltd, Malvern, UK). This instrument has a maximum range of 10km and measures profiles of backscatter and Doppler velocity every 32s, at 36m resolution, and operated continuously between September 2006 and January 2008 when the data presented here was collected. The lidar points directly at vertical: such a configuration is necessary to measure the fall speeds of the ice particles without contamination from the horizontal wind; however it can also lead to strong specular reflections from oriented planar ice crystals, which may bias the Doppler velocities. This is particularly a problem in mid-level, mixed-phase clouds (Westbrook \textit{et al} 2009). Comparison of the backscatter with a second lidar pointing slightly off zenith allows these regions of cloud to be identified and removed - see Westbrook \textit{et al} for more detail. Liquid cloud, profiles with precipitation at the ground, and boundary layer aerosol are also removed using the same procedure described in that paper. This processing effectively limits our sample to stratiform ice clouds which are not raining/snowing at the ground, and our statistics should reflect the properties of these clouds. A total of 1.1~million 32s$\times$36m pixels were analysed: this breaks down into $\approx85000$ pixels per $2.5^{\circ}\mathrm{C}$ interval at $-10^{\circ}\mathrm{C}$, decreasing to $\approx22000$ at $-40^{\circ}\mathrm{C}$.

For lidar measurements at non-absorbing wavelengths, an ice particle with a given shape and orientation is expected to produce a backscatter proportional to its projected area $A$, and the corresponding Doppler velocity would represent the area-weighted fall speed of the crystal population. However, at $1.5\mu\mathrm{m}$ there is some absorption as the light reflects around the inside of the ice crystal: as the particles become larger, the amount of absorption increases, reducing the backscatter. This leads to the lidar Doppler velocity being more strongly weighted towards the smaller particles than a simple area-weighting. We can write this as:
\begin{equation}
\langle v\rangle=\frac{\int N(D)Afv\mathrm{d}D}{\int N(D)Af\mathrm{d}D}+w
\label{velocityweight}
\end{equation}
where $w$ is the vertical wind speed, and the dimensionless factor $f(D)$ falls off as a function of crystal size from $f\approx1$ for small crystals,  to $f\approx0$ for very large ones. As a guide, a backscattered ray of light with a path length of $200\mu\mathrm{m}$ through the ice crystal gives $f=0.45$ (Westbrook \textit{et al} 2009). 

Strictly, this analysis is only correct if the particle is much larger than the wavelength: for the small crystal mode this approximation may not be entirely correct. Mie calculations for ice spheres at $1.5\mu\mathrm{m}$ (not shown for brevity) follow the predicted steady roll off in backscatter efficiency due to absorption for large particles. For very small spheres there is an increasing amount of oscillatory behaviour, leading to a slight enhancement in backscatter efficiency at $25\mu\mathrm{m}$ (the peak of the small mode), and a suppressed efficiency at $15\mu\mathrm{m}$. In practice, for a cloud containing an ensemble of different particle shapes, much of this oscillatory structure will be smoothed out, and it seems unlikely that these departures from the simple geometric optics approximation will have a significant effect on $\langle v\rangle$, or on our conclusions below.

\begin{figure*}
  \noindent\includegraphics[width=39pc]{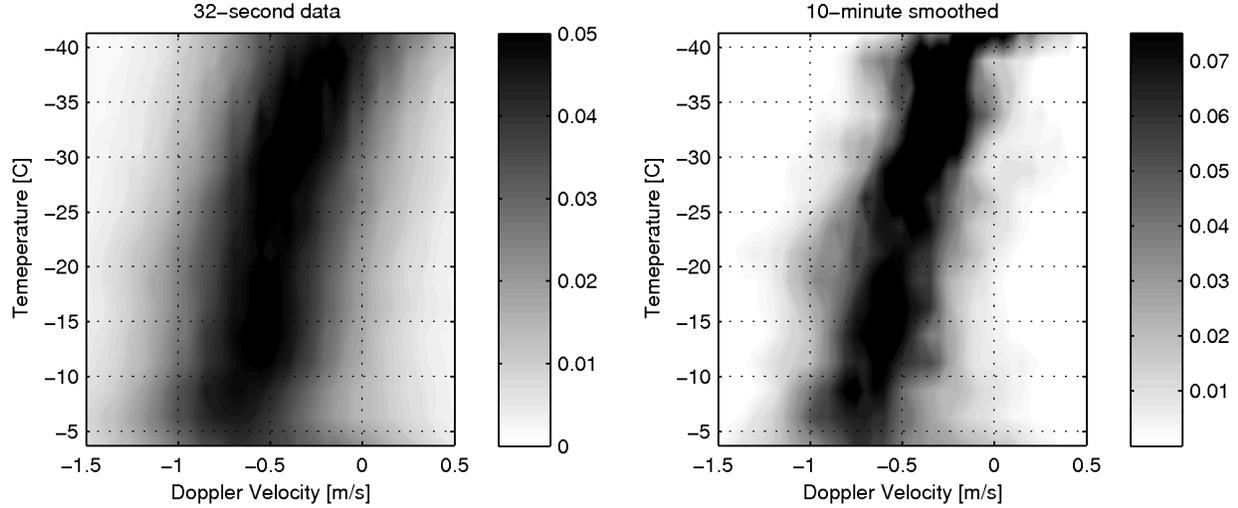}
  \caption{\label{doppler} Filled contour plot showing distribution of lidar Doppler velocity. Left panel shows distribution of raw 32-s data; Right panel shows same when 10~minute smoothing is applied. Grey scale indicates frequency in a particular velocity and temperature bin; frequencies are normalised such that the total in each temperature bin $=1$. Bins are $0.05\mathrm{ms^{-1}}\times 2.5^{\circ}\mathrm{C}$. }
  \end{figure*}

Figure \ref{doppler} shows the distribution of Doppler velocity for all the ice cloud sampled by the lidar over the 17 month period as a function of temperature ($T$ is taken from the Met Office 12~km forecast model output over Chilbolton, see Illingworth \textit{et al} 2007). The distribution is broad, and has a clear temperature trend with particles falling faster at warmer temperatures, indicating the influence of particle growth/aggregation. Note the sign convention of negative Doppler velocities for particles falling toward the lidar. 

Smoothing the lidar data over 10 minute periods before calculating the statistics leads to a narrower distribution as shown in the right hand panel of figure \ref{doppler}, removing small scale variations in $w$, although some influence from lower frequency waves may still be present in the distribution. Since we are looking at stratiform clouds we expect that large scale ascents will be weak. Westbrook (2008) estimated that the maximum fall speed for a compact $60\mu\mathrm{m}$ ice particle is $\approx0.06\mathrm{ms^{-1}}$, so given the observed velocity data we conclude that the small crystal mode does not completely dominate the lidar signal in most clouds; however there could be a more subtle influence. We investigate this by comparing the observed mean values to the forward modelled velocities from Ivanova \textit{et al}'s two-mode size spectra.

\section{Comparison with ice parametrizations}

\subsection{Including small crystals}
\begin{figure*}
  \noindent\includegraphics[width=39pc]{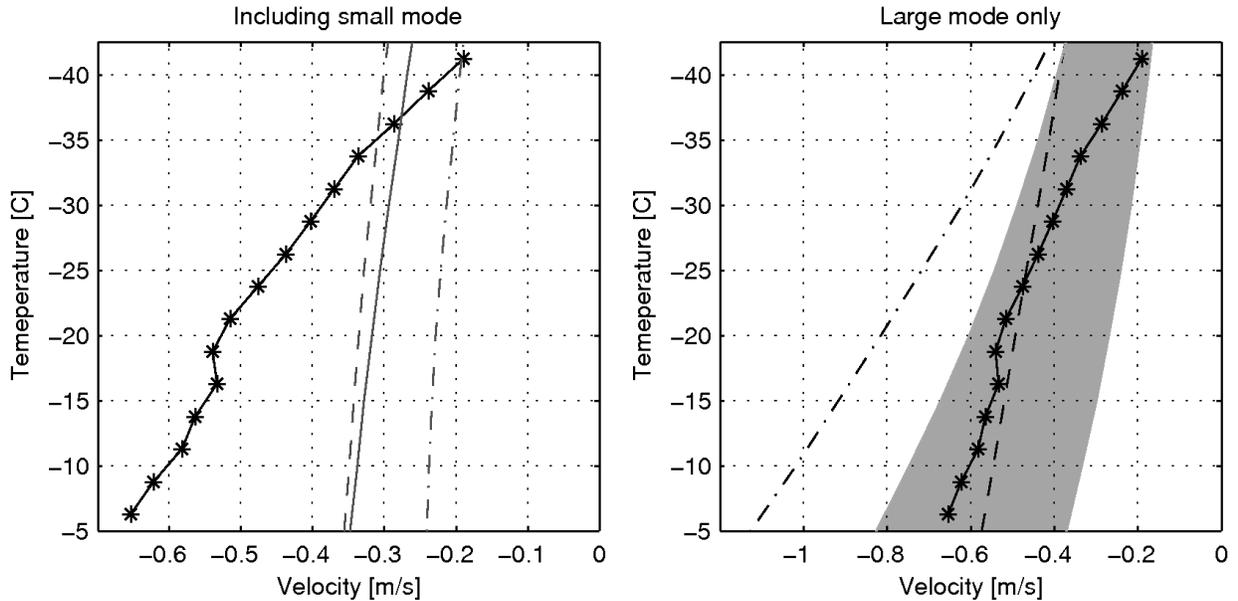}
\caption{\label{meanvt}Average lidar Doppler velocity as a function of temperature (black line with markers). Left panel shows comparison to forward modelled velocities from Ivanova \textit{et al} including the small mode: solid grey line is prediction when the particles are assumed to be planar polycrystals; dashed grey line shows the sensitivity of changing this assumption to aggregates; dot-dash shows change to rosettes. Right panel shows forward modelling of large mode only (black dash). Results of using $\lambda(T)$ relationship measured by Platt (1997) are shown by dot-dash curve; grey shading shows prediction from Wilson and Ballard's scheme for IWCs in the range $0.001-0.1\mathrm{gm^{-3}}$.}
\end{figure*}

We have calculated the mean Doppler velocity for all ice clouds sampled as a function of temperature from the distribution above, and this is plotted in figure \ref{meanvt}. As noted above, the temperature trend is marked, with velocities increasing from $\approx0.2\mathrm{ms^{-1}}$ at $-40^{\circ}\mathrm{C}$ to $\approx0.6\mathrm{ms^{-1}}$ at $-10^{\circ}\mathrm{C}$. For this average over 17 months of data, we expect any residual influence of $w$ in the average values to be positive and very small compared to the total observed Doppler velocities in figure \ref{meanvt}.  

Ice particle size spectra were calculated as a function of temperature as described in Ivanova \textit{et al} (2001), superimposing two gamma distributions:
\begin{equation}
N(D)=N_{0s} D^{\mu_s}\exp(-\lambda_sD)+N_{0\ell} D^{\mu_{\ell}}\exp(-\lambda_{\ell}D).
\label{ivanova}
\end{equation}
For the large particles $\mu_{\ell}=2.64$ and $\lambda_{\ell}$ is diagnosed from $T$ using fits to the aircraft data. For the small mode, $\mu_s=3.24$ and $\lambda_s$ is fixed at $3.65\times10^4\mathrm{m^{-1}}$ (giving a mode centred at approximately $25\mu\mathrm{m}$). There is a simple partitioning of IWC between the two modes: this depends weakly on $\lambda_{\ell}$, and was inferred from aircraft data. For the range of temperatures sampled here, the fraction of the total IWC contained in the small mode was essentially constant at 11\%. This percentage fixes the ratio $N_{0s}/N_{0\ell}$, and the concentration of small crystals relative to the large mode. The spectra in figure \ref{bimodal} are examples of this prescription. 

These size spectra were then used with the mass- and area-diameter relationships assumed in that paper to calculate the area-weighted fall speed of the crystals using the formula in Mitchell (1996). Although Westbrook (2008) shows that Mitchell's formula can overestimate the velocity of small particles, switching to his formula for the small mode had a minimal effect in this case. Air densities were assumed for each temperature based on a standard mid-latitude atmosphere (International Organization for Standardization 1975). Because $\lambda$ (rather than $N_0$) is diagnosed from $T$, the forward modelled velocity for a given temperature does not depend on IWC. 

The forward modelled velocity is plotted as a function of temperature in figure \ref{meanvt}: given the analysis in section 2 we expect this to be faster than the measured lidar velocities due to the absorption effect at $1.5\mu\mathrm{m}$. However for temperatures warmer than $\approx-35^{\circ}\mathrm{C}$ we find the opposite trend: forward modelled velocities are slower than the observations, by almost a factor of two at warm temperatures. We also note that the forward modelled curve has a much weaker temperature trend than the observations, varying from approximately $0.3\mathrm{ms^{-1}}$ at $-40^{\circ}\mathrm{C}$ to $0.35\mathrm{ms^{-1}}$ at $-5^{\circ}\mathrm{C}$. These discrepancies both suggest the small crystal mode is having too strong an influence on the forward-modelled velocity.

One concern about the above analysis is that it may hinge on the assumption of a particular set of (perhaps unrepresentative) mass- and area-diameter relationships. In their parametrization, Ivanova \textit{et al} assume the particles to be planar polycrystals. However they also consider the influence of changing this assumption to bullet-rosettes (leading to a slightly altered partitioning of IWC). Using the relationships in Mitchell (1996) we have forward modelled the result of this alternative assumption, and this is plotted in figure \ref{meanvt}. Now the forward modelling leads to even slower velocities, in the region $0.2\mathrm{ms^{-1}}$; again the temperature trend is much weaker than observed. Finally, it could be argued that in many clouds the large particle mode is typically dominated by aggregates: we have investigated the effect of this, again using the relationships given by Mitchell (1996), and assuming the same partitioning of IWC as for planar polycrystals. The forward modelled fall speeds are very similar to the planar-polycrystal curve. Since analysis of aircraft images (Korolev \textit{et al} 2000) indicates that most ice particles in stratiform clouds are irregular polycrystals or aggregates these three curves should give a good idea of the sensitivity to realistic changes in particle shape, and so we can be reasonably confident that the inconsistency of the forward model with the observations is due to the representation of the particle size spectrum, and not the representation of particle shape.

Given figure \ref{meanvt} it seems clear that the small mode exerts too much influence on the parametrized size spectra for temperatures warmer than $\approx-35^{\circ}\mathrm{C}$, and this adds weight to the suggestion that the measured concentrations of small crystals are likely to be artefacts most of the time. To investigate this further, we have also forward modelled the size spectra with the small mode removed.

\subsection{Large mode only}
Figure \ref{meanvt} shows results of forward modelling Ivanova \textit{et al}'s size spectra if only the large mode is included (dashed line). The velocities are faster, and at warm temperatures are close to the Doppler lidar observations, whilst falling somewhat faster than the lidar velocity at cold temperatures. Also shown is the result of altering the $\lambda(T)$ relationship to that measured by Platt (1997) in midlatitude ice cloud: this leads to somewhat faster fall speeds, lying to the left of the lidar observations at all temperatures; the slope of the curve with temperature is similar to the observations.

For comparison we have also forward modelled velocities from the parametrization of Wilson and Ballard (1998). In this scheme a simple exponential size distribution (equation \ref{gamma} with $\mu=0$) is assumed, with $N_0$ diagnosed from temperature and $\lambda$ deduced from the model IWC. Mass and velocity-diameter relationships are provided in Wilson and Ballard, but no $A(D)$ relationship is defined, so we apply the relationship for aggregates/polycrystals given in Mitchell (1996) to calculate the area-weighting. The grey shaded region in figure \ref{meanvt} shows the range of velocities predicted for IWCs between $0.001$--$0.1\mathrm{gm^{-3}}$, and this velocity range is consistent with the lidar observations. The temperature trend also appears to be realistic given that higher IWCs are correlated with warmer temperatures.

At present it is not possible to evaluate which of these parametrizations is best: if it becomes possible to parameterise $f(D)$ in a sufficiently robust manner, we could forward model the true lidar Doppler velocity (rather than the area weighted velocity), and such a comparison would place a stronger bound on which relationship is most realistic.


\section{Discussion}
Doppler lidar observations have been used to infer whether measurements of high concentrations of small crystals in ice clouds are likely to be genuine or not. Frequency distributions of measured Doppler velocity show a systematic increase in fall speed with temperature, due to broadening of the size distribution. Forward modelling of the area-weighted fall speed from two-mode size-spectra is not consistent with the lidar data and strongly suggest that the large numbers of small ice particles measured in-situ are likely to be instrument artefacts in many cases. Forward modelling with the large mode only gave results which were much more consistent with the observations. We therefore argue that inclusion of a small mode in numerical weather and climate model parametrizations is likely to be unrealistic. However we are cautious not to polarise the issue: clearly in a number of situations small ice particles \textit{are} present, especially near cloud top where there may well be recently frozen droplets or haze particles present, and crystals in the early stages of growth. It would also be valuable to make similar measurements in cold tropical clouds where one might expect high concentration of crystals to be nucleated. However, we conclude that the measurements of numerous small particles at \textit{all heights} in mid-latitude ice cloud are inconsistent with our lidar observations, which are of course unaffected by shattering.

One limitation of this study is that the lidar is unable to penetrate optically thick cloud: this means that in deep systems, only the lowest 2km or so will be observed. This may mean that at cold temperatures the data is biased towards thinner clouds producing less ice (which the lidar can see through), and presumably lower fall speeds. This would imply that the average lidar velocity at cold temperatures may in fact be even faster than in figure \ref{meanvt} and this would act to further reinforce our conclusion about the small crystal mode. Another limitation of this work is that 25\% of the ice cloud sampled was rejected because of specular reflection from oriented planar crystals, predominantly in mid-level clouds. One might worry that this could bias the sample; on the other hand, co-incident radar reflectivity measurements show a similar distribution for both specular and non-specular cloud (Westbrook \textit{et al} 2009). More investigation of the potential influence of this removed cloud on our statistics would be valuable.

Future work will focus on calculation of $f(D)$ using ray tracing simulations of polycrystal and aggregate shapes; this will then allow us to completely forward model the true lidar velocities from model parametrizations. We also note that implicit in equation \ref{velocityweight} is the assumption that the particle shape (or ensemble of different shapes) does not change systematically as a function of particle size. Since we have taken a large sample of different clouds we might expect any such correlation to average out; theoretical sensitivity studies would be valuable to back this up. 

For models which diagnose $N_0$ as a function of $T$ and predict $\lambda$ from IWC, coincident radar reflectivity ($Z$) measurements may allow the parametrizations to be tested, since $Z$ is related to $\lambda$ through $N_0$. The combination of simultaneous radar reflectivity, radar Doppler spectra and lidar Doppler velocity has the potential both to test model parametrizations as a whole, and also to help discriminate which individual elements of that parametrization are realistic and which are not. We aim to persue this approach in future work to evaluate the Met Office Unified Model cloud scheme.


%
%
%
%
%
%

%
%
%
%

\begin{acknowledgments}
We would like to thank the staff at the Chilbolton Facility for Atmospheric and Radio Research for operation and maintainance of the Doppler lidar. This work was carried out under the Natural and Environmental Research Council grants NER/Z/S/2003/00643 and NE/EO11241/1.
\end{acknowledgments}

%
%
%
%
%
%
%
%
%
%


%
%

\end{article}




%
%
%
%
%
%


\end{document}